# Delineating the magnetic field line escape pattern and stickiness in a poloidally diverted tokamak

Caroline G. L. Martins[1], M. Roberto[1,#] and I. L. Caldas[2]

[1]Departamento de Física, Instituto Tecnológico de Aeronáutica, São José dos Campos, São Paulo, 12228-900 Brazil

[2]Universidade de São Paulo; Instituto de Física 05315-970 São Paulo, SP, Brazil

E-mails: carolinegameiro@gmail.com, marisar@ita.br and ibere@if.usp.br

[#]Correspondent author

**ABSTRACT**

We analyze a Hamiltonian model with five wire loops that delineates the magnetic surfaces of the tokamak ITER, including a similar safety factor profile and the X-point related to the presence of a poloidal divertor. Non-axisymmetric magnetic perturbations are added by external coils, similar to the correction coils installed at the tokamak DIII-D and those that will be installed at ITER. To show the influence of magnetic perturbations on the field line escape, we integrate numerically the field line differential equations and obtain the footprints and deposition patterns on the divertor plate. Moreover, we show that the homoclinic tangle describes the deposition patterns in the divertor plate, agreeing with results observed in sophisticated simulation codes. Additionally, we show that while chaotic lines escape to the divertor plates, some of them are trapped, for many toroidal turns, in complex structures around magnetic islands, embedded in the chaotic region, giving rise to the so called stickiness effect characteristic of chaotic Hamiltonian systems. Finally, we introduce a random collisional term to the field line mapping to investigate stickiness alterations due to particle collisions. Within this model, we conclude that, even reduced by collisions, stickiness still influences the field line transport.

I. Introduction

In conventional tokamaks the plasma column is separated from the wall by a physical limiter made of a material that resists the impact and the temperature of plasma particles [1]. An alternative separation of the plasma column and the chamber's wall can be achieved installing poloidal divertors.

Divertors are essential components in modern tokamaks, such as ITER [2], and they consist of conductors arranged externally, that carry electric currents in the same direction of the plasma current, in the toroidal direction of the tokamak. A X point (or hyperbolic fixed point) will appear at the positions where the poloidal magnetic field is null, due to the overlap of the magnetic fields of the conductors with the magnetic field of the plasma. From the X point arises a separatrix with two manifolds, one stable and another unstable. Outside the separatrix the surfaces intersect the collector plates, which play a role similar to the physical limiter in conventional tokamaks [3, 4].

Numerical equilibria reconstructions to simulate plasmas in the presence of poloidal divertors are well known in the literature [5]. However, the computation to integrate magnetic field lines, for such magnetic configuration, is time-consuming [6-8]. On the other hand, simple models can delineate quite well dynamical properties of open and closed field lines near the separatrix. In particular, a simple way to delineate the MHD equilibrium magnetic field in tokamaks is to consider the magnetic configuration created by a set of coils or wires [9, 10].

Wires carrying electric current generate concentric circles of magnetic field lines around themselves. If another wire is positioned parallel to the first one, magnetic field lines are no longer circular, although they are still closed curves around each wire. For such magnetic configuration, there will be a point where the magnetic field is zero. In this position arise two manifolds of the separatrix, forming a X point. A large number of wires distort the magnetic fields, creating more than one X point and their respective separatrixes.

Non-axisymmetric magnetic perturbations destroy the magnetic separatrix creating homoclinic tangles, leading to the formation of a layer of chaotic field lines [5, 11]. Due to the flexibility of models based on magnetic fields produced by wires, one can consider different effects by adequately specifying the perturbations. Thus, perturbed wire models have been used to analyze a range of dynamical properties in double-null (two X-points) and single-null (one X-point) diverted tokamaks, such as, the width of the scrape-off layer [12], chaotic layer formation [13-15] and particle drift orbits [16].

A model described by five parallel infinite wires was analyzed in [9], to simulate the magnetic surfaces of the tokamak ITER, in the presence of magnetic perturbations created by error fields due to asymmetries on the external coils [13, 17]. Moreover, a Hamiltonian with three loop wires was analyzed in [10], to describe two mapping methods and to study the stochastic field lines in poloidal divertor tokamak plasmas affected by external non-axisymmetric magnetic perturbations.

Following reference [10], in the present work we analyze a Hamiltonian model with five loop wires that enables the choice of magnetic axis position, triangularity and elongation. The versatility of the model allows us to delineate equilibrium magnetic surfaces that reproduce ITER like magnetic topology, including a similar safety factor profile. Here, we add to equilibrium perturbations created by pairs of loop coils carrying opposite flowing currents, introduced in [10]. Those perturbations are similar to the ones related to the correction coils (C-coils) installed at the DIII-D tokamak [5] and those that will be installed at ITER [18].

Thus, using the mentioned Hamiltonian description we solve numerically the perturbed magnetic field line differential equations and show the influence of magnetic perturbations on the deposition patterns at the divertor plate. To do that, we calculate the numbers of toroidal turns, called connection lengths, performed by the perturbed magnetic field lines until reaching the divertor plates and their non uniform distribution, the deposition patterns, on the divertor plates [5, 6, 11]. Moreover, we show that the homoclinic tangle describes the deposition patterns in the divertor plate, agreeing with results observed in sophisticated simulation codes [19, 20]. Additionally, we show that even so the chaotic lines escape to the divertor plates, some of them are trapped, for many toroidal turns, in complex structures around magnetic islands, embedded in the chaotic region, giving rise to the so called stickiness effect characteristic of Hamiltonian systems [21]. So, our results indicate that the deposition on the plates depends on the sticky structure of the analyzed magnetic configuration. Finally, we introduce a random collisional term to the field line mapping to reproduce stickiness alterations due to particle collisions. Within this model, we conclude that, even reduced by collisions, stickiness still influences the field line transport.

In section II we introduce the equilibrium magnetic surfaces of our five loop wires model. In section III we present the resonant perturbation caused by external loop coils. Numerical results on homoclinic tangle formation and examples of escape pattern typical of those computed for tokamaks are presented in section IV. Stickiness effect investigations are in section V, and its alterations due to particle collisions in section VI; finally, in section VII we present the conclusions.

**II. Equilibrium Model**

To describe the equilibrium magnetic field lines, we choose a coordinate system appropriated to the tokamak symmetry, shown in figure 1.

**FIGURE 1**

We will consider in this work a simple model that consists of five wire loops carrying electric currents, with the same relative position proposed in [9], as following,

| n | $R_n$ (m) | $Z_n$ (m) | $I_n$ (MA) |
|---|---|---|---|
| 1 (plasma) | 6.41 | 0.513 | 15.00 |
| 2 | 3.72 | -7.580 | 15.90 |
| 3 | 3.20 | 8.600 | 16.28 |
| 4 | 2.45 | 0.513 | -5.69 |
| 5 | 10.00 | 0.513 | -4.60 |

Table 1 - Wires positions and currents values

As introduced in the Appendix A, for a tokamak with large aspect ratio, $(R_0/a \gg 1)$, the unperturbed Hamiltonian, can be approximated to,

$$H_0(z, p_z) = \frac{-1}{R_0} \sum_{j=1}^{5} \frac{\mu_0 I_j}{4\pi B_0} \left( \ln\left( 64 \left[ \left(\frac{R}{R_0} - \frac{R_j}{R_0}\right)^2 + \left(\frac{Z}{R_0} - \frac{Z_j}{R_0}\right)^2 \right]^{-1} \right) - 4 \right) \quad (1)$$

The wire 1 represents the plasma current. The role of the wires 2 and 3 is to create the lower and upper X points, respectively. These two X points are located in distinct separatrices: the active separatrix (internal) and the inactive separatrix (external). The outer separatrix is called inactive because the plasma does not reach this position; on the other hand, it plays an important role in the shape of the surfaces. The negative currents in wires 4 and 5 compress the left and right sides of the magnetic surfaces, then one can model the desired elongation of the plasma column.

**FIGURE 2**

Figure 2(a) shows the positions of the intersections of the five loops with the R-Z plane and the surfaces generated by them. Figure 2(b) shows the surfaces related to the plasma column and two X points: the upper X point of the external separatrix (inactive), and the lower X point of the internal separatrix (active), that defines the boundary of the plasma.

Each magnetic surface has a well-defined characteristic known as rotation number, which is the average of the poloidal angle performed by a field line after a full toroidal turn. In this model, the poloidal angle is the angular displacement in the plane R-Z made by a field line. For tokamaks, the inverse of the rotational number is the safety factor, and it is used to characterize the topology of the lines.

In order to obtain the safety factor profile of the magnetic surfaces, we integrate the unperturbed magnetic field lines equations (see Appendix A) for many initial conditions located on a horizontal line, indicated in figure 2(b). We consider that a field line completes $m$ toroidal turns if $\varphi = 2m\pi R_0$. Thus, on rational surfaces with safety factor $q = m/n$, the periodic field lines perform $m$ toroidal turns and $n$ poloidal turns.

**FIGURE 3**

Figure 3 shows the safety factor profile calculated from our model, for initial conditions located at the auxiliary line of figure 2(b). In this figure it is possible to identify the positions of two surfaces with infinite safety factor. These surfaces correspond to the two separatrixes shown in Figure 2(b). In a magnetic system, hyperbolic points correspond to the positions with null poloidal magnetic field and, consequently, the values of the safety factor goes to infinity. Profiles similar to that shown in Figure 3 are expected for the tokamak ITER [22].

### III. Perturbed Model by External Coils

Many tokamaks have non-axisymmetric perturbation coils designed specifically to create chaotic layers in the peripheral region of the plasma column [23-25]. Despite this, few theoretical or experimental data may be found to understand the effects of these chaotic layers in plasmas with elongation and triangularity in the presence of poloidal divertors.

**FIGURE 4**

In this work we consider a non-axisymmetric perturbation generated by $N = 10$ pairs of loop coils carrying opposite currents, $\pm I_c$, positioned at $(R_{c+}, Z_{c+}) = (10.2, 3)$ and $(R_{c-}, Z_{c-}) = (10.2, -3)$. The perturbation created by these coils is similar to the correction coils (C-coils) installed at the DIII-D tokamak [5], and those that will be installed at ITER [18]. An illustration of this design can be seen in figure 4, which shows the transversal cross section of the tokamak ITER, and ten pairs of coils arranged around the chamber.

For a large aspect ratio tokamak, the perturbing Hamiltonian, $H_1(z, p_Z, \varphi)$, related to the scheme shown in figure 4, can be approximated by [10],

$$\varepsilon H_1(z, p_Z, \varphi) = \frac{\mu_0 I_c(\varphi)}{4\pi B_0 R_0} \left( \ln\left( \left( \frac{R-R_{c+}}{R_0} \right)^2 + \left( \frac{Z-Z_{c+}}{R_0} \right)^2 \right) - \ln\left( \left( \frac{R-R_{c-}}{R_0} \right)^2 + \left( \frac{Z-Z_{c-}}{R_0} \right)^2 \right) \right) \quad (2)$$

where the current in the coils, $I_c(\varphi)$, is periodically changing along the toroidal angle, $\varphi$, and can be represented by the discontinuity function, $I_c(\varphi) = (-1)^k I_c$, for $(\pi/N)k < \varphi < (\pi/N)(k+1)$ and $k = 0,...,(2N-1)$. This function can be expanded in Fourier series [10]:

$$I_c(\varphi) = \frac{4I_c}{\pi} \sum_{p=0}^{\infty} \frac{\sin[(2p+1)N\varphi]}{2p+1} \tag{3}$$

We will consider only the first term, $p = 0$, of the function (3), since this term gives the main effect of the perturbation. The contribution of higher order toroidal modes, $(2p+1)N$, decreases exponentially while increasing $p$ [10]. Therefore, the perturbing Hamiltonian is reduced to

$$\varepsilon H_1(z, p_Z, \varphi) = \varepsilon \frac{\mu_0 I_p}{\pi^2 B_0 R_0} \left( \ln\left( \left(\frac{R - R_{c+}}{R_0}\right)^2 + \left(\frac{Z - Z_{c+}}{R_0}\right)^2 \right) - \ln\left( \left(\frac{R - R_{c-}}{R_0}\right)^2 + \left(\frac{Z - Z_{c-}}{R_0}\right)^2 \right) \right) \sin[N\varphi] \tag{4}$$

where the perturbation parameter is $\varepsilon = (I_c / I_p)$ [10].

The total Hamiltonian in canonical coordinates of position and momentum is given by,

$$H(z, p_Z, \varphi) = H_0(z, p_Z) + \varepsilon H_1(z, p_Z, \varphi) \tag{5}$$

**FIGURE 5**

The set of equations

$$\begin{aligned} \frac{dz}{d\varphi} &= \frac{\partial H}{\partial p_Z} \\ \frac{dp_Z}{d\varphi} &= -\frac{\partial H}{\partial z} \end{aligned} \tag{6}$$

was numerically integrated with perturbation parameter $\varepsilon \cong 0.0047$ that corresponds to $I_c = 70kA$. The value used for the current in the coils was deducted from reference [5], which shows the range of the current amplitude applied in the correction coils (C-coil) installed at the DIII-D tokamak, which is, approximately, 0.266% to 0.533% of the plasma current. By the same consideration, we define a limit range for the current in our perturbation coils for ITER tokamak, which is therefore, $I_c = 40kA$ to $I_c = 80kA$.

Figures 5(a) and 5(b) show a chaotic layer around the lower hyperbolic point. The magnetic field lines are no longer closed, and, eventually, reach the divertor plates located horizontally at $Z = -3.7m$ (represented by the black segment in figure 5(b)), following the manifolds that leave the X point. One can notice, in figures 5(b), magnetic islands immersed in the chaotic layer, which play an important role in the field lines escape [26].

**IV. Homoclinic Tangle and Escape Patterns**

Since the topology of the manifolds is highly unstable and can be destroyed by arbitrary perturbations [11, 27], it is essential to study the behavior of the field lines near the separatrix and particularly around the X point, to understand the way in which particles are transported through the separatrix [28, 29].

Accordingly, we consider our model to calculate the escape of the field lines to the divertor plates, placed horizontally at $Z = -3.7m$ (black segment in figure 5(b)). As the particles follow the field lines, the structure of escape obtained must be closely related to measurable profiles of particle deposition on the divertor plates [5]. To analyze the escape of the field lines in question, we calculated the connection length, which is the number of toroidal turns performed by a field line until it reaches the plate. The field line is integrated forward and backward in $\varphi$, for initial conditions in a box with $4.88 \leq R_0 \leq 5.02$ and $-3.50 \leq Z_0 \leq -3.35$. Then a color map is constructed where the color indicates the number of toroidal turns that the field line needs to reach the plate, and the axes, $(R, Z)$, represent the initial position of the field lines.

**FIGURE 6**

The manifolds of figure 6(a) were approximated by numerically integrating initial conditions in a small grid constructed around the hyperbolic fixed point (R ≈ 4.9506m, Z ≈ -3.4428m). Figure 6(a) shows the stable (blue) and unstable (red) manifolds of the lower hyperbolic fixed point (in black) for system perturbed by $N = 10$ pairs of loop coils with $I_c = 70kA$. Each individual manifold does not intersect itself, but the two manifolds intersect each other an infinite number of times, forming a complex pattern known as homoclinic tangle [11, 5]. Figure 6(b) shows the connection lengths of the case shown in 6(a), where the region in white color represents the initial position of field lines that take less than one toroidal turn to escape to the divertor plates. It is noticed in figure 6(b) a structure similar to the homoclinic tangle of stable and unstable manifolds shown in figure 6(a). This type of structure has been observed through the use of computer codes that simulate escape patterns in reference [19].

One can calculate the magnetic footprint which is the set of points that reaches the divertor plates. The magnetic field line is integrated backwards in $\varphi$, for initial conditions located at the divertor plate (black segment in figure 5(b)), in a box with $5.065 \leq R_0 \leq 5.080$ and $0 \leq \varphi_0 \leq 2\pi$. Then a color map is constructed where the color indicates the connection length of the field line, and the axes, $(\varphi, R)$, represent the initial position of the field lines.

**FIGURE 7**

The magnetic footprints of figure 7(a) and the zoom in the rectangle in figure 7(b), show structures created by magnetic field lines at the divertor plate (black segment in figure 5(b)). Each color represents the connection length, namely the number of toroidal turns required for the field line to reach the divertor plate. The manifolds of the separatrix act as boundaries for the magnetic footprints, determining their position and shape. Outside the boundaries of the separatrix one can find, in white, magnetic field lines that take less than one toroidal turn to escape to the divertor plate. We note that there are ten lobes related to the number of pairs of loops coils used in our model (see equation (4)). Since the structure of the homoclinic tangle is self-similar, we expect magnetic footprints to have a fractal nature [30]. Similar results were observed in sophisticated simulation codes [28, 20].

**V. Stickiness Effect**

Chaotic field lines escape to the divertor plates, but some of them may be trapped for many toroidal turns (i.e. long connection lengths) in complex structures at the border of magnetic islands, giving rise to the so-called stickiness effect characteristic of Hamiltonian systems [21]. Since the distribution of the connection lengths directly interfere on the particle transport, determining the deposition patterns on the divertor plates [5], it is essential to analyze the topology of sticky structures.

**FIGURE 8**

Figure 8 shows the connection lengths of field lines near the X point, located at the reference line, in green, of figure 9(a), varying according to the amplitude of the current in the perturbation coils. One can notice in figure 8(a) horizontal structures with long connection lengths (red stripes) located between regions with short connection lengths. Some of these horizontal structures (red stripes) remain intact while the current in the coils is increased. Figure 8(b) shows a zoom in the black rectangle of figure 8(a), in order to clarify some details of the horizontal structure (red stripe) located at $-3.080 \leq Z \leq -3.062$.

# FIGURE 9

Figure 9(a) shows the magnetic surfaces perturbed by $N = 10$ pairs of loop coils with $I_c = 70 kA$. The reference line in green was used to calculate the connection lengths distribution shown in figure 8. Figure 9(b) shows a zoom of the red square of (a), constructed with the same height and position of the red stripe shown in figure 8(b), i.e. $-3.080 \leq Z \leq -3.062$. In figure 9(b) one can identify islands formed by field lines located on regular surfaces, and also a concentration of pixels (iterations) on the chaotic layers surrounding the island, that interfere in the field line escape.

# FIGURE 10

Figure 10(a) shows a zoom at the border of the island of figure 9(b). Figure 10(b) shows the rotation number for the field line with initial condition represented by the red dot in (a), located at the chaotic layer surrounding the island. One can notice a plateau in the rotation number for the initial 1500 toroidal turns, indicating that the field line is trapped in a resonance with rotation number $\omega = 51/250$. After that, the field line escapes entering in the chaotic sea, and its rotation number for the next toroidal turns diverges. Figures 10(c) and 10(d) show the Poincaré map of the commented field line for (c) 1500 toroidal turns and (d) 5000 toroidal turns, and the trapping effect become clearer. For the 1500 initial toroidal turns the field line is trapped at the resonance with $\omega = 51/250$ and after that, the field line escapes and, eventually, hits the divertor plate.

Figures 10(b), 10(c) and 10(d) evidence the so-called stickiness effect that should be caused by the presence of cantori, surrounding resonance islands. The cantori gaps are usually very small, thus a chaotic orbit inside them takes a long time before escaping to the outer chaotic sea, and a stickiness phenomenon appears [21]. One can find many stick regions in the analyzed area of figure 8(a), and each one of them is related to different sticky resonances.

**VI. Stickiness Effect in a Collisional Scenario**

Particle transport inside the plasma is not only determined by the magnetic field lines, but also by collisions. So, to estimate the stickiness effect of section V, in a collisional scenario, we have to modify the field line equations to simulate how collisions divert particle center guide from the field line trajectories. It has been verified, recently, in a simple numerical model, that the manifolds of the X points still govern the particle dynamics when collisions are included [31]. To check that, we add the effect of an additional noise in the system that simulates collisional diffusion of particles inside the plasma column. The effect of such noise is represented by adding a vector of random orientation to the magnetic field lines equation, at every complete toroidal turn [32], thus, equations for the magnetic field lines become,

$$\frac{dz}{d\varphi} = \frac{\partial H}{\partial p_z} + \rho \sin(\theta_t)$$
$$\frac{dp_z}{d\varphi} = -\frac{\partial H}{\partial z} + \rho \cos(\theta_t)$$
(7)

where $\rho$ is the collisional amplitude, $0 \leq \theta_t \leq 2\pi$ is a random phase, $(z, p_z)$ are the canonical coordinates of position and momentum, respectively, and the toroidal angle, $\varphi$, the canonical time. We choose the collisional amplitude, $\rho$, by analyzing the mean free path obtained for ITER [9].

**FIGURE 11**

Figure 11(a) shows the sticky island from figure 9(b) but in a collisional regime with collisional amplitude $\rho = 1 \times 10^{-3}$. One can notice the destruction of the internal structures that used to form the island, including the structures that used to form the border of the island. Figure 11(b) shows the rotation number for the field line with initial condition represented by the red dot in 10(a), and the stickiness effect did not vanish with the addition of the collisional noise. Although, instead of being trapped for 1500 toroidal turns, the field line was trapped for 500 toroidal turns. It suggests that collisions, depending on their amplitude, do not extinct stickiness structures, but decrease their time scale of trapping.

## VII. Conclusions

The Hamiltonian model presented in this work is capable of reproducing magnetic surfaces similar to those expected in ITER equilibrium, as well as similar profiles of safety factor. An external perturbation, similar to the perturbation of the C-coils installed at the tokamak DIII-D and those that will be installed at ITER, was added to study qualitatively the dynamical characteristics of the magnetic field lines in a chaotic divertor layer. By tracing the manifolds we showed the influence of the homoclinic tangle on the deposition patterns of field lines in the divertor plates, agreeing qualitatively with results obtained with sophisticated computer codes. The stickiness effect caused by resonances embedded in the chaotic region was analyzed, indicating that this effect survives for long ranges of current in the perturbation coils, trapping magnetic field lines for many toroidal turns. A random noise was added to the field line equation to simulate collisions between the particles, and the stickiness effect did not vanish under the presence of such collisions. However, the time scale of the trapping decreased when compared to the case without collisions.

Acknowledgments: The authors would like to thank Dr. T. Kroetz for useful comments, and the following Brazilian scientific agencies for the financial support: CAPES, CNPq and São Paulo Research Foundation (FAPESP), Grants 2010/13162-0, 2013/03401-6 and 2011/19269-11.

**REFERENCES**
[1] W. M. Jr. Stacey, *Fusion Plasma Analysis* (John Wiley & Sons, New York, 1981)
[2] ITER Physics Expert Group on Divertor, Nucl. Fusion **39**, 2391 (1999)
[3] F. WAGNER, et al., Phys. Rev. Lett. **49 (19)**, 1408 (1982)
[4] M. Kikuchi, K. Lackner and M. Tran, *Fusion Physics* (IAEA, Vienna, 2012)
[5] T. E. Evans, R. A. Moyer and P. Monat, Phys. Plasmas **9**, 4957 (2002)
[6] A. Wingen, T. E. Evans and K. H. Spatschek, Nucl. Fusion **49**, 055027 (2009)
[7] T. A. Casper, W. H. Meyer, L. D. Pearlstein and A. Portone, Fus. Eng. and Design **83**, 552 (2007)
[8] M. Brix, N. C. Hawkes, A. Boboc, V. Drozdov, S. E. Sharapov and JET-EFDA Contributors, Rev. Sci. Instrum. **79**, 10F325 (2008)
[9] T. Kroetz, Caroline G. L. Martins, M. Roberto and I L. Caldas, J. Plasma Phys. **79** (05), 751 (2103)
[10] S. S. Abdullaev, K. H. Finken, M. Jakubowski and M. Lehnen, Nucl. Fusion **46**, S113 (2006)
[11] E. C. da Silva, I. L. Caldas, R. L. Viana and M. A. F. Sanjuan, Phys. Plasmas **9** (12), 4917 (2002)
[12] A. H. Boozer and A. B. Rechester, Phys. Fluids **21**, 662 (1978)
[13] N. Pomphrey and A. Reiman, Phys. Fluids B **4**, 938 (1992)
[14] H. Ali, A. Punjabi, A. Boozer and T. E. Evans, Phys. Plasmas **11**, 1908 (2004)
[15] H. Ali, A. Punjabi and A. Boozer, J. Plasma Phys **75**, 303 (2008)
[16] U. Daybelge and C. Yarim, J. Nucl. Mat. **266-269**, 809 (1999)
[17] A. Reiman, Phys. Plasmas **3**, 906 (1996)
[18] A. Foussat, P. Libeyre, N. Mitchell, Y. Gribov, C. T. J. Jong, D. Bessette, R. Gallix, P. Bauer, and A. Sahu, IEEE Trans. Appl. Supercond. **20 (3)**, 402 (2010)
[19] O. Schmitz et al., Plasma Phys. Control. Fusion **50**, 124029 (2008)
[20] A. Wingen, T. E. Evans and K. H. Spatschek, Phys. of Plasmas **16**, 042504 (2009)
[21] G. Contopoulos and M. Harsoula, Int. J. Bifurcation Chaos **20**, 2005 (2010)

[22] G. Janeschitz, *The Status of ITER; The ITER Design Review* (presented to APS-DPP Town Meeting on ITER Design Review, Orlando, FL, 2007)

[23] K. H. Finken, S. S. Abdullaev, T. Eich, D. W. Faulconer, M. Kobayashi, R. Koch, G. Mank, and A. Rogister, Nucl. Fusion **41**, 503 (2001)

[24] Ph. Ghendrih, A. Grosman, and H. Capes, Plasma Phys. Controlled Fusion **38**, 1653 (1996)

[25] M. Z. Tokar, Phys. Plasmas **6**, 2808 (1999)

[26] T. Kroetz, M. Roberto, E. C. da Silva, I. L. Caldas and R. L. Viana, Phys. Plasmas **15**, 092310 (2008)

[27] R. K. W. Roeder, B. I. Rapoport and T. E. Evans, Phys. Plasmas **10**, 3796 (2003)

[28] M. W. Jakubowski et al., Nucl. Fusion **49 (9)**, 095013 (2009)

[29] M. W. Jakubowski, S. S. Abdullaev, K. H. Finken, M. Lehnen and the TEXTOR Team, J. Nucl. Mater **532**, 337 (2005)

[30] R. L. Viana, E. C. da Silva, T. Kroetz, I. L. Caldas, M. Roberto and M. A. F. Sanjuan, Phil. Trans. R. Soc. A **369**, 371 (2010)

[31] A. B. Schelin, I. L. Caldas, R. L. Viana and M. S. Benkadda, Phys. Lett. A **376**, 24 (2011)

[32] P. Beaufume, M. A. Dubois and M. S. Benkadda, Phys. Lett. A **147**, 87 (1990)

## Appendix A – Unperturbed Hamiltonian Equation

In this appendix we present the unperturbed Hamiltonian equation, in $(R, Z, \varphi)$ coordinates (see figure 1), introduced in [10], and used in the present article. Accordingly, the magnetic field line equations are:

$$\frac{1}{R}\left(\frac{dZ}{d\varphi}\right) = \frac{B_z}{B_\varphi}$$
$$\frac{1}{R}\left(\frac{dR}{d\varphi}\right) = \frac{B_R}{B_\varphi} \quad (1)$$

The magnetic field $\vec{B}$ can be expressed by the vector potential $\vec{A}(R, Z, \varphi) = (A_R, A_Z, A_\varphi)$, so that $\vec{B} = \nabla \times \vec{A}$. Therefore, the component of the magnetic field can be written as,

$$B_R = \frac{1}{R}\left(\frac{\partial A_Z}{\partial \varphi}\right) - \left(\frac{\partial A_\varphi}{\partial Z}\right)$$
$$B_\varphi = -\frac{\partial A_Z}{\partial R} \quad (2)$$
$$B_Z = \frac{1}{R}\left(\frac{\partial RA_\varphi}{\partial R}\right)$$

We introduce canonical variables $(z, p_Z)$ associated with the geometric coordinates $(R, Z)$ and the magnetic field, according to [10],

$$z = \frac{Z}{R_0}$$
$$p_Z = \frac{1}{B_0 R_0} \int_{R_0}^{R} B_\varphi dR \qquad (3)$$

The toroidal magnetic field is expressed by $B_\varphi(R) = B_0 R_0 / R$, then $Z = R_0 z$ and $R = R_0 e^{p_z}$, and the equations for the field lines can be transformed to Hamiltonian form,

$$\frac{dz}{d\varphi} = \frac{\partial H}{\partial p_Z}$$
$$\frac{dp_Z}{d\varphi} = -\frac{\partial H}{\partial z} \qquad (4)$$

The variables $(z, p_Z)$ are the canonical coordinates of position and momentum, respectively, and the toroidal angle, $\varphi$, the canonical time.

The vector potential for each current loop is given by [10],

$$A_\varphi(R, Z) = \sum_{j=1}^{5} \frac{\mu_0 I_j}{\pi k_j} \sqrt{\frac{R_0}{R}} \left[ \left(1 - \frac{k_j^2}{2}\right) K(k_j) - E(k_j) \right] \qquad (5)$$

where $K(k_j)$ and $E(k_j)$ are the complete elliptic integrals with module,

$$k_j^2 = \frac{4 R_0 R}{(R + R_0)^2 + (Z - Z_j)^2}, \quad j = 1 \ldots 5. \qquad (6)$$

For a tokamak with large aspect ratio, $(R_0 / a \gg 1)$, the unperturbed Hamiltonian, $H_0(z, p_Z) = -(R A_\varphi(R, Z))/(B_0 R_0^2)$, can be approximated to,

$$H_0(z, p_Z) = \frac{-1}{R_0} \sum_{j=1}^{5} \frac{\mu_0 I_j}{4\pi B_0} \left( \ln\left( 64 \left[ \left(\frac{R}{R_0} - \frac{R_j}{R_0}\right)^2 + \left(\frac{Z}{R_0} - \frac{Z_j}{R_0}\right)^2 \right]^{-1} \right) - 4 \right) \qquad (7)$$

**FIGURES AND CAPTIONS**

**FIGURE 1**

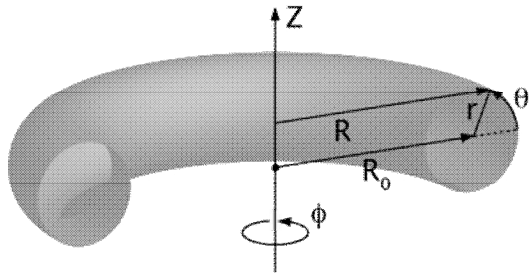

Figure 1 - Geometry of a toroidal system

**FIGURE 2**

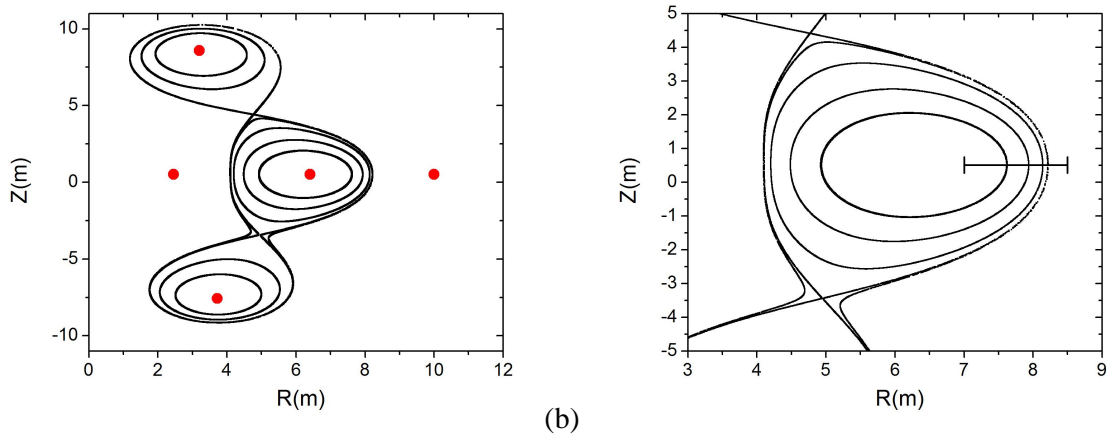

(a)                                          (b)

Figure 2 - (a) Magnetic surfaces. The red dots represent the intersection of the five loops with the plane R-Z. (b) Zoom in (a), showing the plasma column, and the upper and lower X points, forming the outer separatrix (inactive) and the inner separatrix (active), respectively. The auxiliary line indicates the initial conditions to calculate the safety factor of figure 3.

**FIGURE 3**

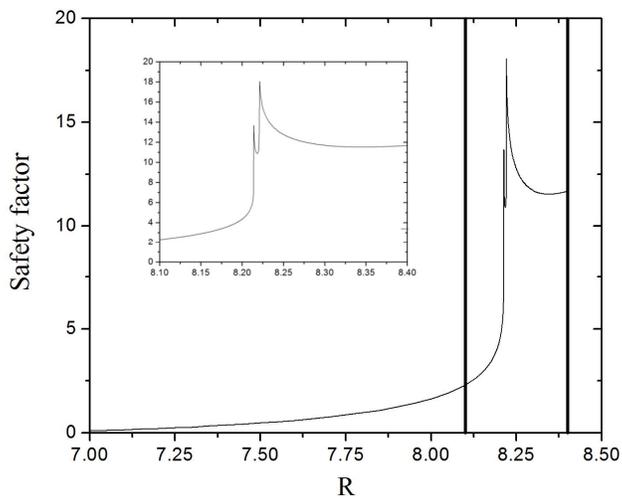

Figure 3 - Safety factor profile for initial conditions located at the auxiliary line from figure 2(b). The rectangle indicates the region amplified. The two points where the safety factor values go to infinity represent the two separatrixes formed by the lower and upper X points.

**FIGURE 4**

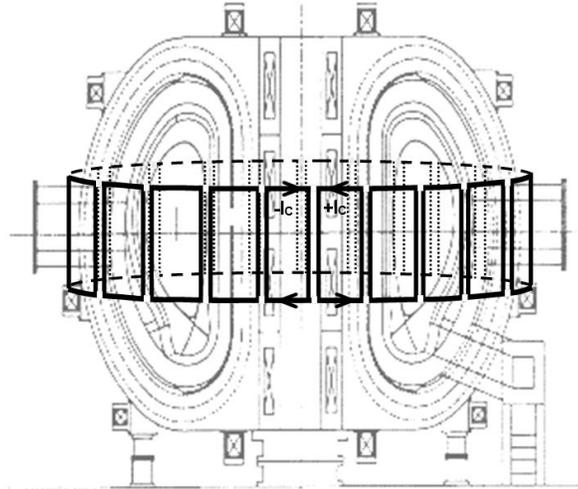

Figure 4 - Transversal cross section of the tokamak ITER with the addition of $N=10$ pairs of perturbation coils located on the equatorial plane of the chamber, carrying opposing currents, $\pm I_c$, at positions $(R_{c+}, Z_{c+}) = (10.2, 3)$ and $(R_{c-}, Z_{c-}) = (10.2, -3)$.

**FIGURE 5**

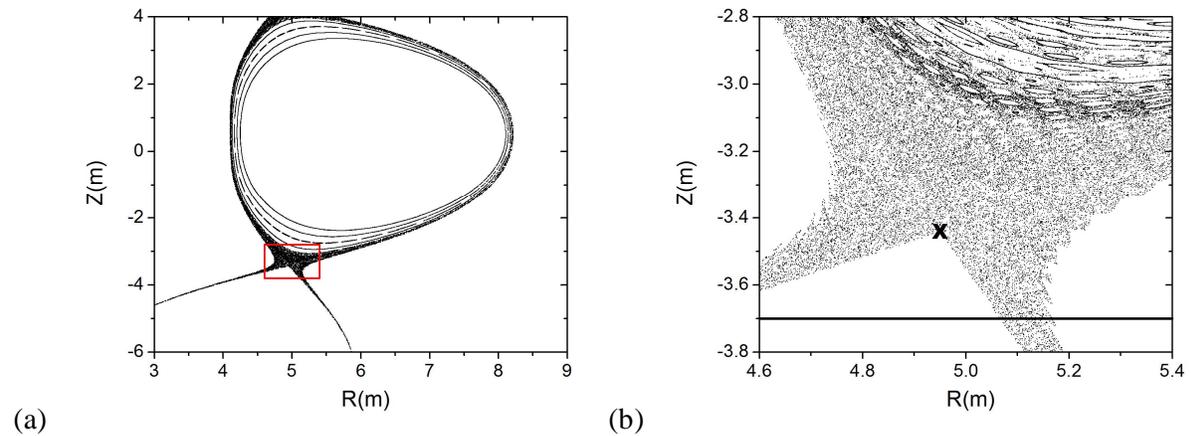

(a) (b)

Figure 5 - (a) Magnetic surfaces perturbed by $N=10$ pairs of loop coils with $I_c = 70kA$. (b) Zoom in (a) showing some islands immersed in the chaotic region. The solid lined segment represents the divertor plate. The black X shows the position of the hyperbolic fixed point.

**FIGURE 6**

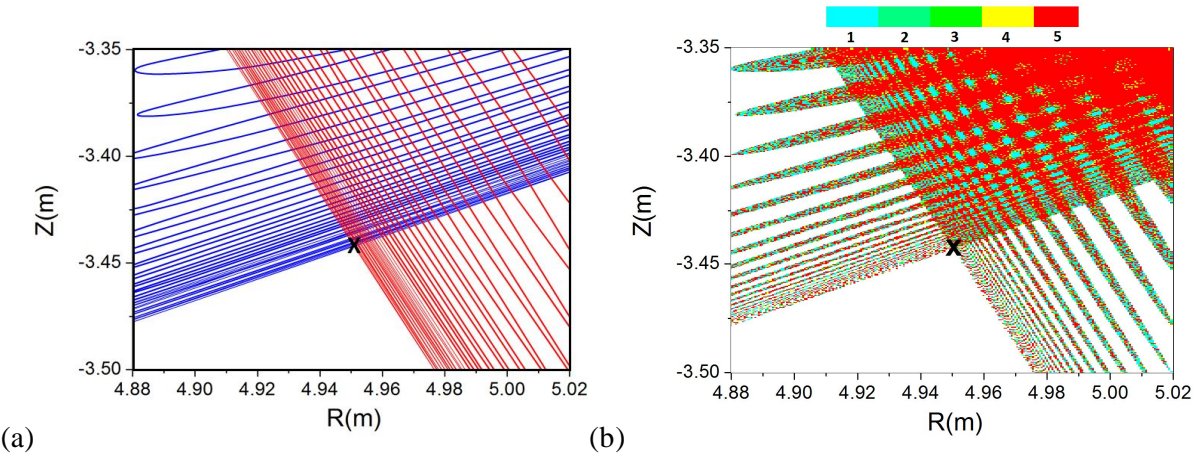

Figure 6 – Numerical calculations for system perturbed by $N = 10$ pairs of loop coils with $I_c = 70 kA$ (a) Homoclinic tangle formed by the unstable (red) and stable (blue) manifolds from the lower X point (in black) (b) Connection lengths of the homoclinic tangle in the region near the X point (in black).

**FIGURE 7**

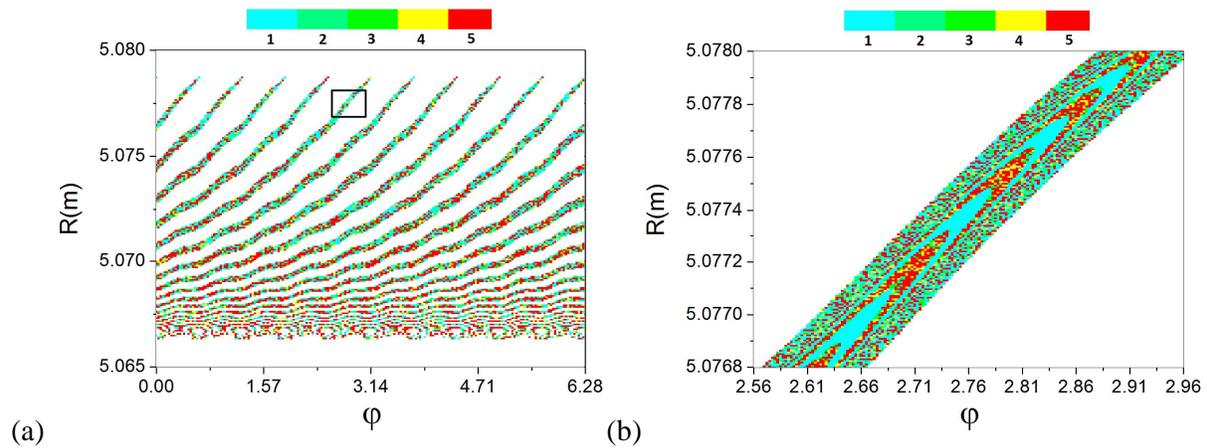

Figure 7 - (a) Magnetic footprints for the system perturbed by $N = 10$ pairs of loop coils with $I_c = 70 kA$, showing places at the divertor plate where magnetic field lines began their trajectories, and their correspondent connection lengths. (b) Zoom at the rectangle shown in (a), emphasizing details of the structure formed by the homoclinic tangle.

**FIGURE 8**

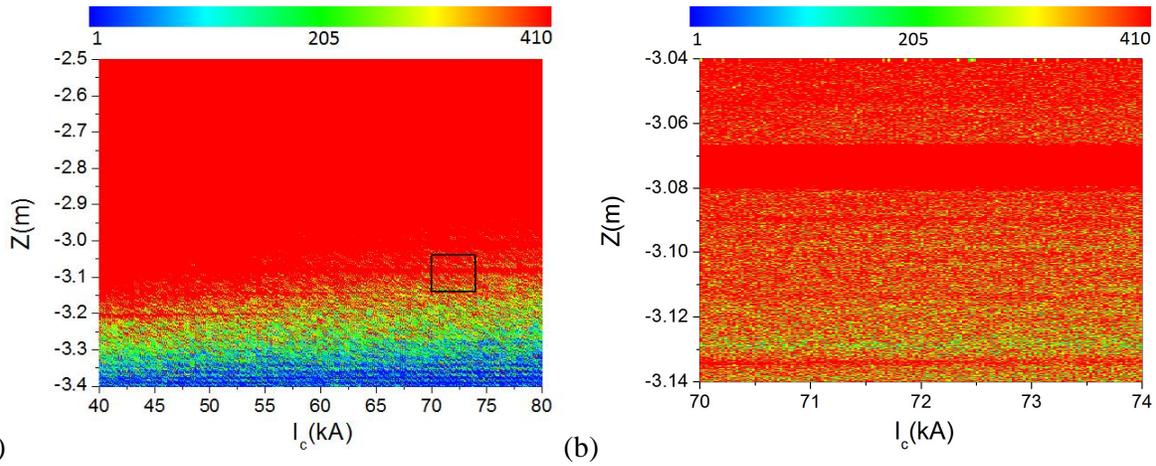

Figure 8 - (a) Color map of the connection lengths as a function of $I_c$ and $Z(m)$, showing the existence of horizontal structures (red stripes) that remain intact for long ranges of current in the coils. (b) Zoom at the black rectangle of (a), emphasizing details of a horizontal structure (red stripe) at $-3.080 \leq Z \leq -3.062$.

**FIGURE 9**

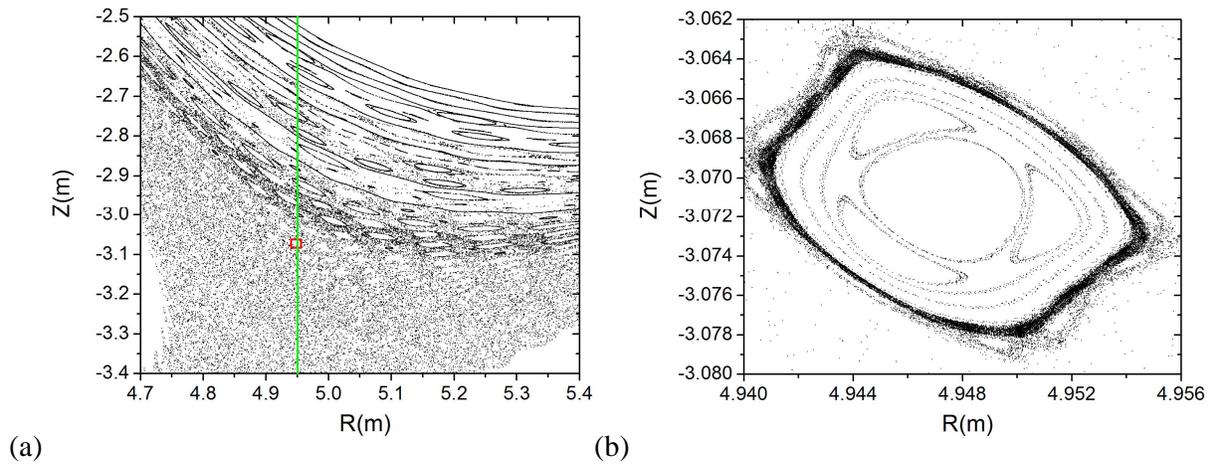

Figure 9 - (a) Magnetic surfaces perturbed by $N = 10$ pairs of loop coils with $I_c = 70 kA$. The reference line in green was used to calculate the connection lengths distribution shown in figure 8 (b) Zoom at the red rectangle shown in (a), emphasizing details of the island that causes the red stripe shown in figure 8(b) at $-3.080 \leq Z \leq -3.062$.

**FIGURE 10**

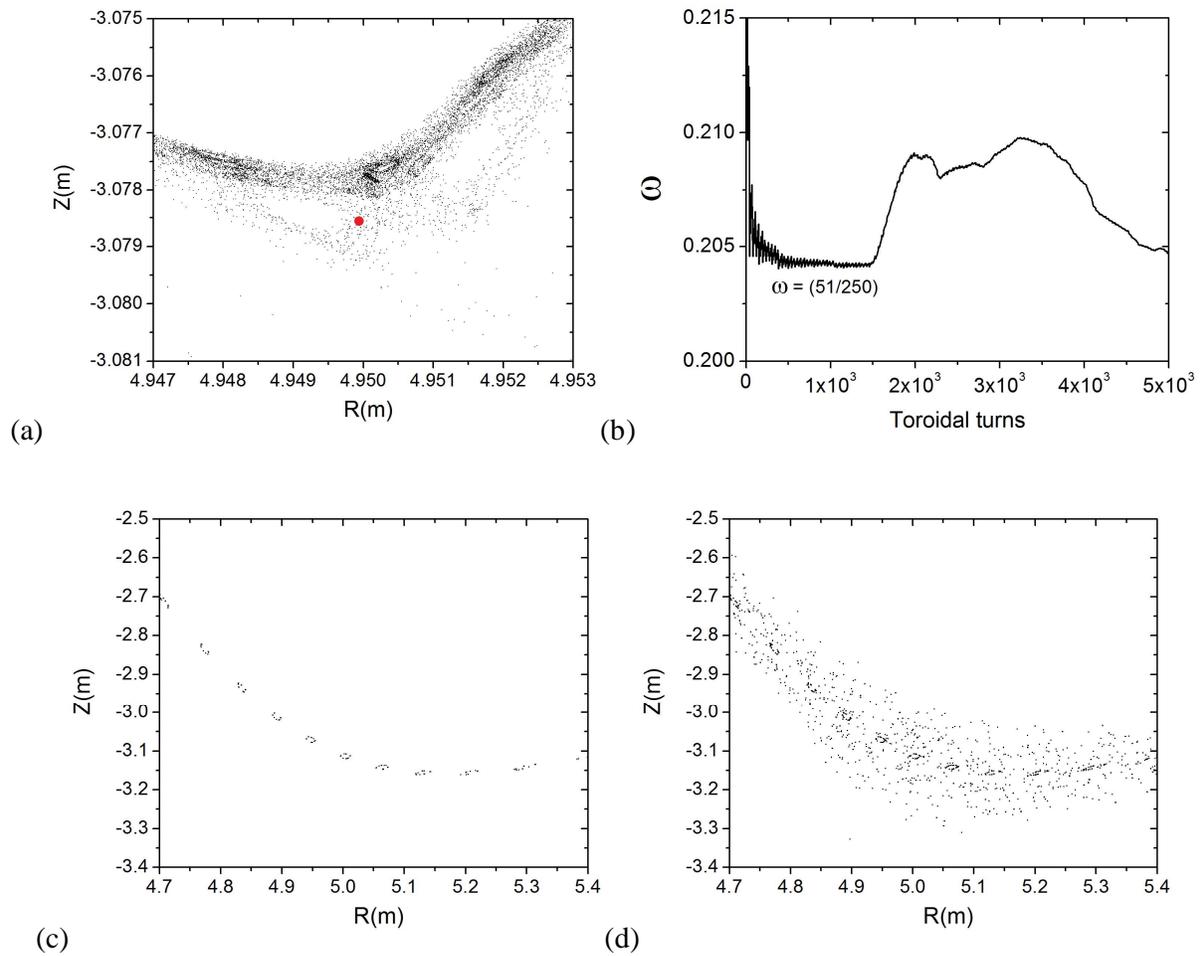

Figure 10 - (a) Zoom at the border of the island of figure 9(b) (b) Rotation number of the initial condition represented by the red dot in (a) (c) Initial condition at the red dot in (a) iterated for 1500 toroidal turns (d) Initial condition at the red dot in (a) iterated for 5000 toroidal turns.

**FIGURE 10**

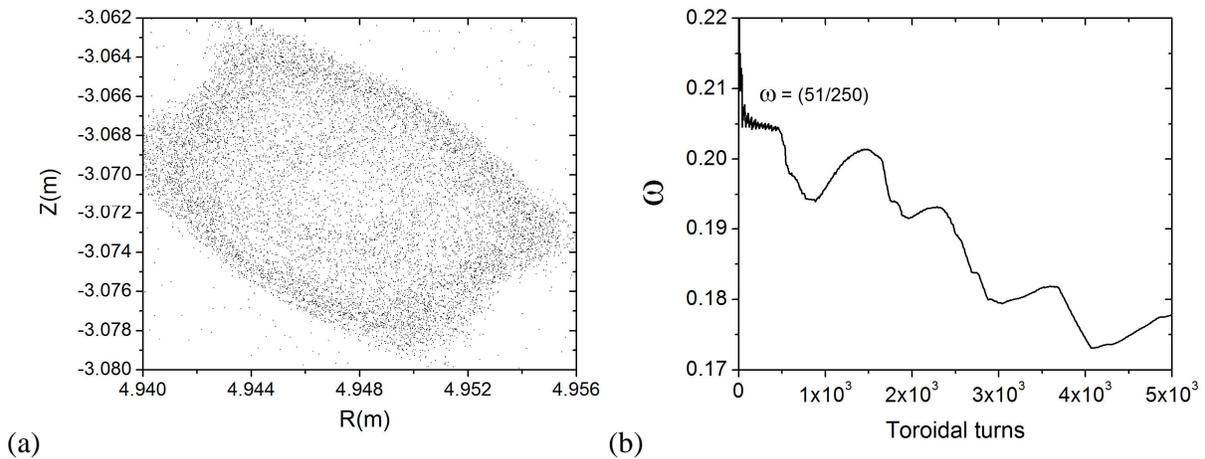

Figure 11 - (a) Sticky island from figure 9(b) in a collisional regime with $\rho = 1\times 10^{-3}$. (b) Rotation number of the initial condition represented by the red dot in figure 10(a) in a collisional regime.